# Transparent Dense Sodium

Yanming Ma[1,2], Mikhail Eremets[3], Artem R. Oganov[2,4*], Yu Xie[1], Ivan Trojan[3,5], Sergey Medvedev[3], Andriy O. Lyakhov[2*], Mario Valle[6], and Vitali Prakapenka[7]

[1] *National Lab of Superhard Materials, Jilin University, Changchun 130012, P. R. China*
[2] *Laboratory of Crystallography, Department of Materials, ETH Zurich, Wolfgang-Pauli-Str. 10, CH-8093 Zurich, Switzerland*
[3] *Max Planck Institute für Chemie, Postfach 3060, 55020 Mainz, Germany*
[4] *Geology Department, Moscow State University, 119992 Moscow, Russia*
[5] *on leave from A.V. Shubnikov Institute of Crystallography, RAS, 117333, Leninskii pr.59, Moscow, Russia*
[6] *Data Analysis and Visualization Services, Swiss National Supercomputing Centre (CSCS), Cantonale Galleria 2, 6928 Manno, Switzerland*
[7] *Consortium for Advanced Radiation Sources, University of Chicago, Chicago, Illinois 60637, USA*

*\*Now at Department of Geosciences and New York Center for Computational Science, Stony Brook University, Stony Brook, NY 11794-2100, USA.*

**Under pressure, interatomic distances in materials decrease and the widths of electronic valence and conduction bands are expected to increase, eventually leading to metallization of all materials at sufficiently strong compression; however, core electrons can also substantially overlap at densities achieved with current high-pressure techniques, dramatically altering the electronic properties predicted for archetypes of the free-electron metals lithium (Li)[1-3] and sodium (Na)[4,5] and creating Peierls-like distortions, leading in turn to structurally complex phases[6-8] and superconductivity with a high critical temperature[9-11]. The most intriguing prediction, of the insulating states due to the pairing of alkali atoms under pressure in Li[1] and Na[4], has yet to be experimentally confirmed. Here we show that at pressures of ~200 GPa (~5.0 fold compression), Na becomes**

**optically transparent with the formation of a wide gap dielectric, not in the paired state, but in the monatomic state. This insulating state is formed due to p-d hybridizations of valence electrons and their repulsion by core electrons into the interstices of the six-coordinated highly distorted double-hexagonal close-packed structure. It is possible that such insulating states with localised interstitial electrons play an important role in the physics of strongly compressed matter.**

Sodium adopts the body-centered cubic (b.c.c.) structure at ambient conditions. Under pressure, it transforms to face-centered cubic (f.c.c.) structure at 65 GPa[12] and to cI16 structure at 103 GPa[7,13]. On further compression up to 160 GPa, a number of phases have recently been discovered in a narrow pressure–temperature range near the minimum of the melting curve ($T_{melt} \approx 300$ K at 118 GPa)[7,8,13]. At pressures above 190 GPa, theoretical calculations[4,14] suggest the stability of the body-centered tetragonal (Cs-IV) and α-Ga (*Cmca*) structures. The α-Ga structure is especially interesting because pairing of Na atoms results in a zero band gap above 800 GPa[4]; however, because these earlier calculations were based on educated guesses of likely structures, there exists the possibility that hitherto unsuspected structures are stable rather than α-Ga, thereby changing our understanding of the metal–insulator transition. In this context, we undertook an extensive experimental and theoretical study into insulating states in Na at high pressures.

Our X-ray diffraction (XRD) data at 101 and 113 GPa reveal the known f.c.c. ($a = 3.46$ Å) and cI16 phases ($a = 5.44$ Å), respectively, in agreement with available experimental data[7,12,15]. The Raman spectra show no features at pressures below 130 GPa, however, a pronounced Raman spectrum appeared at higher pressures (Fig. 1a), indicating a major phase transformation associated with a gradual decrease in the reflection of visible light from the sample. The Raman spectra are in good accordance with the theoretical spectra for the experimentally observed[7,8] oP8 (*Pnma*) phase (Fig. S1d). Above 150 GPa, the Raman spectra show a marked change, including a strong decrease in intensity, signifying another phase transition (Fig. 1a). The XRD pattern of this phase is consistent with the tI19 structure[7,8].

Remarkably, Na becomes optically transparent at pressures of ≈ 200 GPa (208 GPa in one run (Fig. 2a) and 194 GPa in another (Fig. S1a). The band gap of the transparent Na appears to be at least ~1.3 eV, as measured from the edge in the absorption spectrum (Fig. S1c). The onset of transparency coincides with dramatic changes in Raman spectra (Fig. 1a), with the appearance of a single intense line at ~340 cm$^{-1}$. On releasing the pressure, the transparent phase exists to 182 GPa, at which point the sample reverted to opaque and yielded the Raman spectra of the tI19 phase (Fig. 1a). On further decompression, Na transforms to the oP8 phase at 132 GPa; finally, at 120 GPa, the sample reverted to a highly reflective metallic state, devoid of any Raman signal.

We have performed extensive structure searches, unbiased by any prior knowledge and based on global optimisation using the *ab initio* evolutionary methodology for crystal structure prediction[16-19]. Our variable-cell simulations yielded the experimentally known f.c.c.[7,12], cI16[7,8,15], and oP8[7,8] phases below 250 GPa. In the pressure range of 152–260 GPa, the enthalpies of the oP8 and tI19 phases (Fig. 3) are highly competitive and the energy landscape of Na is complex. Our simulated structure search identified many distinct structures having enthalpies within 30 meV/atom of the ground state.

At higher pressures of 320 and 1000 GPa, simulations revealed a simple but unusual structure, which can be described as double-hexagonal close-packed (d.h.c.p.) structure squeezed along the *c*-axis (Pearson symbol hP4). This structure has an extremely small *c/a* ratio (e.g., 1.391 at 320 GPa), less than half the ideal value in normal d.h.c.p structure, which characterises the densest packing of spherical particles, $c/a = 2(8/3)^{1/2} = 3.266$. Atoms in this structure have a six-fold coordination (Fig. 4a). The characteristic double-well energy profile (Fig. S8a) and the six-coordinate topology mean that Na-hP4 can be considered as a new structure type, being structurally, energetically, and electronically distinct from normal d.h.c.p. Remarkably, the energy barrier (Fig. S8) between two minima in Na-hP4 occurs at $c/a = 3.266$, where Cs-d.h.c.p. has a stable minimum.

Of note, the electronic structure of Na-hP4 (Fig. 4b) reveals an intriguing insulating state characterised by a large occupancy of d-orbitals involved in p-d hybridizations. GW calculations (known to provide accurate description of band gaps, within 10% of experimental values[20]) show (Fig. 4d) that Na-hP4 possesses a large energy gap of 1.3 eV at 200 GPa (7.88 Å$^3$/atom) and the gap rapidly increases with increasing pressure and reaches 6.5 eV at 600 GPa (5.03 Å$^3$/atom).

An analysis of experimental and theoretical data enables us to identify the observed transparent phase as Na-hP4. The optical transparency in dense Na is in agreement with large values of band gap found theoretically. The quality of the XRD pattern obtained for the transparent phase is insufficient to derive an exact structure solution; however, the observed reflections are well indexed as Na-hP4, with lattice parameters of $a = b = 2.92$ Å and $c = 4.27$ Å at 190 GPa (Fig. 2b). The position of the observed Raman peak and its dependence on pressure (Fig. 1b) are in excellent agreement with theoretical calculations for Na-hP4, which possesses only one Raman-active $E_{2g}$ mode. Enthalpy calculations (Fig. 3) confirmed the stability of Na-hP4 beyond the tI19 phase. Note the existence of a clear discrepancy between the predicted tI19→hP4 transition pressure of 260 GPa (Fig. 3) and the observed pressure (≈200 GPa). While this discrepancy may in part be due to the athermal nature (i.e., the neglect of entropic contributions) of our calculations, we expect that the main reason is the well-known overstabilisation of metallic states by DFT calculations (e.g., Ref.[19]).

Our calculations provide insight into the physical origin of the insulating state. Upon compression, $3d$ bands show a rapid drop in energy relative to the $3p$ bands, and increasingly hybridize with them[4]. This hybridization is the key to strong electron localisation: a marked charge accumulation occurs only in the open interstitial regions (Figs. 4c, S6, and S7). The shortest Na–Na distance in Na-hP4 is 1.89 Å at 300 GPa (6.67 Å$^3$/atom), which implies strong core–valence overlap (the $3s$ and $2p$ orbital radii in Na are 1.71 and 0.28 Å, respectively) or even a significant core–core overlap (the ionic radius of Na$^+$ is 1.02 Å) between neighboring Na atoms. Considering the geometry of charge maxima relative to the atoms, we can propose pd and pd$^2$ hybridizations for Na1 in the octahedral sites and Na2 in the triangular prismatic coordination, respectively, consistent with the site-*l*-projected density of states (Figs. 4b and S10). Thus, the insulating state in Na-hP4 is explained by the strong localisation of valence electrons (Figs. 4c and S6) in the interstices of Na-hP4. While the stacking of close-packed layers of Na atoms is C<u>A</u>C<u>B</u>C<u>A</u>C<u>B</u>… (Fig. 4a), as in any d.h.c.p. structure, the interstitial charge density maxima are located only in layers A and B (i.e., in every other layer), and form a nearly-perfect h.c.p arrangement (layer stacking of <u>ABAB</u>… and *c/a* ~ 1.3–1.6, close to the ideal *c/a* = 1.633 for the h.c.p. structure). This finding explains the approximate halving of the *c/a* ratio as a reflection of the close packing of the interstitial electron density maxima, rather than Na atoms. This suggests that repulsion between the interstitial electron pairs is a major structure-forming interaction in Na-hP4. In contrast to the p-d hybridization in Na, our calculations[3] for other alkalis suggest they are dominated either by p-electrons (Li) or by d-electrons (K, Rb, and Cs). Interestingly, Na-hP4 might be viewed as the intermediate structure between the six-coordinated structure predicted for Li and the ideal d.h.c.p. structures of K, Rb, and Cs[3].

Since the number of ionic cores (each stripped of one valence electron) is exactly twice that of interstitial electron density maxima, Na-hP4 is an analogue of Ni$_2$In-type structure (considered also in Ref. 7), with ionic cores sitting at the Ni-sublattice, while the interstitial density maxima at the In-sublattice. Pressure-induced transitions both in the A$_2$X compounds[28] and in Na lead to an increase of the coordination number of X sites (i.e. holes – rather than atoms in Na structures) – these coordination numbers are 8 in cI16, 9 in oP8, and 11 in hP4 phases of Na.

The case of dense Na provides an unexpectedly extreme example of a wide-gap insulator created by compression of a "simple" metal. This novel dielectric is formed by ionic cores and localised interstitial electron pairs. There exists a strong analogy with electrides; i.e., compounds in which the interstitial electron density maxima play the role of anions. Such insulating states may be favoured in other elements and compounds when atomic cores strongly overlap – e.g. in planetary and stellar interiors, where they might be detectable via their low electrical and thermal conductivities.

**METHODS SUMMARY**

Our diamond-anvil cell (DAC) experiments employed diamonds with 30–50 μm culet beveled to 8–9° and rhenium, tungsten, or cubic BN powder mixed with epoxy gaskets. Sodium of 99.95% purity (Alfa Aesar) was loaded in a glove box in an atmosphere of pure nitrogen containing <0.1 ppm oxygen and water. Prior to loading, the DAC was kept at 130°C in a vacuum for 20 hours to remove traces of water and gases at the surface of the gasket, thereby excluding reaction of the clamped sodium with rhenium or cBN/epoxy gaskets. Pressure was measured mainly by the Raman shift of the high-frequency edge of the stressed diamonds[22], and was in good agreement with the pressure obtained from X-ray diffraction data for Re and cBN in checking experiments. We estimate the pressure error to be within 5 GPa at the highest pressures, and less at lower pressures. An extremely sharp Raman edge of the diamond anvil indicated an excellent hydrostaticity of Na up to the highest pressures. An HeNe 25 mW power laser was used for excitation of Raman spectra. Angle dispersive X-ray diffraction studies were performed at station 13-IDD at the Advanced Photon Source, Argonne National Laboratory, USA ($\lambda = 0.3344$ Å).

The evolutionary structure search was performed using the USPEX code[16-18], and the underlying *ab initio* structure relaxations were performed using DFT within the generalised gradient approximation (GGA)[23] and the frozen-core all-electron projector-augmented wave (PAW)[24,25] method, as implemented in the VASP code[26]. The quasiparticle energies were calculated within the GW approximation, as implemented in the ABINIT code[27], in which self-energy corrections are added to the Kohn–Sham energies at the selected *k* points that characterise the DFT band gap.

**Acknowledgements** We thank the Swiss National Science Foundation (grants 200021-111847/1 and 200021-116219), CSCS, and ETH Zurich for the use of supercomputers. We acknowledge partial support from DFG grants Er 539/1/2-1 and China 973 Program (No. 2005CB724400). Part of this work was performed at **Author Contributions** Y.M. proposed the research and predicted the new structures. Y.M., Y.X., and A.R.O. did the calculations. M.E., I.T., S.M., and V.P. performed the experiments. Y.M., A.R.O., and M.E. conducted the data analysis, generated most of the ideas, and wrote the paper. A.L. wrote the latest version of the structure prediction code, and M.V. helped in data analysis. Y. M, M.E., and A.R.O contributed equally to this paper.

**Author information** The authors declare no competing financial interests. Correspondence and requests for materials should be addressed to Y.M. (mym@jlu.edu.cn).

## Figure Captions

**Figure 1 | Raman spectra of sodium. a,** Raman spectra obtained at increasing pressures. Spectra of cI16, oP8, tI19, and transparent phases, where the intensities of different phases are representative, are shown by green, black, blue, and red lines, respectively. **b,** Pressure dependence of Raman peaks for five pressure runs. The colours correspond to the spectra in **a**. Filled and empty symbols correspond to runs on pressure increase and decrease, respectively. Black lines are theoretical Raman modes for the oP8 structure (thicker lines for intensive peaks); the red line is the $E_{2g}$ mode for transparent Na-hP4. No Raman calculations were performed for the complex tI19 phase.

**Figure 2 | Phase transformations in Na at megabar pressures. a,** Photographs of

the Na sample taken under combined transmitted and reflected illumination (see also Fig. S1) at releasing pressure for different phases: 199 GPa (transparent phase), 156 GPa (tI19), 124 GPa (oP8), and 120 GPa (cI16). The transparent phase was created at 208 GPa, but remained when pressure was dropped to 199 GPa, similar to the outcome of another run (Fig. S1). **b,** X-ray CCD image and integrated diffraction pattern taken at 190 GPa from the transparent Na sample. Tick marks show the calculated reflection positions for Na-hP4. Asterisks indicate reflections from Re gasket.

**Figure 3 | Enthalpy curves (relative to f.c.c.) as a function of pressure for cI16, CsIV, α-Ga, oP8, tI19, and hP4 structures**. The inset shows the transition sequence f.c.c. → cI16 → tI19 (oP8) → hP4 in more detail. Two phase transitions, cI16 → tI19 → hP4, are predicted at 152 and 260 GPa, respectively. The oP8 structure is less stable than the tI19 phase in the pressure range of 152–260 GPa, but is energetically very competitive. It is noteworthy that temperature effects (or lattice energies) are not considered in the calculation, but might play an important structural role. Note also that tI19 is an incommensurate structure; thus, its enthalpy calculations were performed with the periodic approximant (Fig. S4). The continuous merging of the enthalpy curves of oP8 and hP4 at 250 GPa is due to the second-order nature of the oP8→hP4 transition (Fig. S2).

**Figure 4 | Structural and electronic properties of Na-hP4. a,** Crystal structure of Na-hP4 (space group $P6_3/mmc$). Lattice parameters at 320 GPa (6.50 $Å^3$/atom): $a = b = 2.784$ Å and $c = 3.873$ Å; two inequivalent atomic positions: Na1 at $2a$ (0.0, 0.0, 0.0)

and Na2 at 2*d* (2/3, 1/3, 1/4). **b**, Band structure (left panel) and partial densities of states (right panel) at 300 GPa (6.67 Å$^3$/atom). **c**, Difference charge density (e/Å$^3$) (crystal density minus superposition of isolated atomic densities) plotted in the (110) plane at 320 GPa (6.50 Å$^3$/atom). **d,** Band gaps calculated by DFT with the PAW potentials using VASP code (solid circles) and the pseudopotential method using ABINIT code (solid squares), and the GW band gap (solid triangles) as a function of volumes for Na-hP4. The upper axis represents theoretical pressure.

**METHODS**

We treated the 2*s*, 2*p*, and 3*s* electrons as valence in the PAW potential adapted from VASP library and used the plane-wave kinetic energy cutoff of 910 eV, which gave excellent convergence of the total energies, energy differences, and structural parameters. We used the Monkhorst–Pack k meshes of 16×16×16, 12×12×12, 19×19×13, 16×16×12, 12×16×12, 6×6×6, and 20×20×14 for f.c.c., cI16, α-Ga, Cs-IV, oP8, tI19, and hP4, respectively, which gave excellent convergence of the total energies (within 1 meV/atom) in the enthalpy calculations. GW calculations for Na-hP4 were performed with 131 *k* points in the first BZ, and an energy cutoff of 20 Ry was chosen for calculation of the Coulomb matrix. In the calculation of self-energy matrix, 18 occupied bands and 120 unoccupied bands were explicitly treated. With the choice of these parameters, the band gaps were found to converge within 0.01 eV.